\documentclass[12pt]{article}
\usepackage{amsfonts,epsf}

\def\hybrid{\topmargin 0pt      \oddsidemargin 0pt
	\headheight 0pt \headsep 0pt
	\textwidth 6.25in       
        \textheight 9.5in       
	\marginparwidth .875in
	\parskip 5pt plus 1pt   \jot = 1.5ex}

\catcode`\@=11
\def\marginnote#1{}

\newcount\hour
\newcount\minute
\newtoks\amorpm
\hour=\time\divide\hour by60
\minute=\time{\multiply\hour by60 \global\advance\minute by-\hour}
\edef\standardtime{{\ifnum\hour<12 \global\amorpm={am}%
	\else\global\amorpm={pm}\advance\hour by-12 \fi
	\ifnum\hour=0 \hour=12 \fi
	\number\hour:\ifnum\minute<10 0\fi\number\minute\the\amorpm}}
\edef\militarytime{\number\hour:\ifnum\minute<10 0\fi\number\minute}

\def\draftlabel#1{{\@bsphack\if@filesw {\let\thepage\relax
   \xdef\@gtempa{\write\@auxout{\string
      \newlabel{#1}{{\@currentlabel}{\thepage}}}}}\@gtempa
   \if@nobreak \ifvmode\nobreak\fi\fi\fi\@esphack}
	\gdef\@eqnlabel{#1}}
\def\@eqnlabel{}
\def\@vacuum{}
\def\draftmarginnote#1{\marginpar{\raggedright\scriptsize\tt#1}}

\def\draft{\oddsidemargin -.5truein
	\def\@oddfoot{\sl preliminary draft \hfil
	\rm\thepage\hfil\sl\today\quad\militarytime}
	\let\@evenfoot\@oddfoot \overfullrule 3pt
	\let\label=\draftlabel
	\let\marginnote=\draftmarginnote
   \def\@eqnnum{(\theequation)\rlap{\kern\marginparsep\tt\@eqnlabel}%
\global\let\@eqnlabel\@vacuum}  }

\def\numberbysection{\@addtoreset{equation}{section}
	\def\theequation{\thesection.\arabic{equation}}}

\catcode`@=12
\relax

\def\nn{\nonumber}
\def\beq{\begin{equation}}
\def\eeq{\end{equation}}
\def\bea{\begin{eqnarray}}
\def\eea{\end{eqnarray}}

\def\e{{\rm e}}
\def\N{{\mathbb N}}
\relax
\numberbysection
\hybrid

\begin{document}
\begin{titlepage}
\begin{center}
{\large\bf Classification of Conformal Field Theories Based on Coulomb
gases. Application to Loop Models.}\\[.3in] 
        {\bf Vladimir S. Dotsenko (1), Jesper Lykke Jacobsen (2) and Marco
Picco (1)} \\  
	{\bf (1)} {\it LPTHE\/}\footnote{Unit\'e Mixte de Recherche CNRS
UMR 7589.},\\
        {\it  Universit{\'e} Pierre et Marie Curie, PARIS VI\\
              Universit{\'e} Denis Diderot, PARIS VII\\
	      Bo\^{\i}te 126, Tour 16, 1$^{\it er}$ {\'e}tage \\
	      4 place Jussieu,
	      F-75252 Paris CEDEX 05, FRANCE}\\
              and\\
	{\bf (2)} {\it Laboratoire de Physique Th\'eorique et Mod\`eles
                   Statistiques, Universit\'e Paris-Sud, \\
                   B\^atiment 100, 91405 Orsay, France.}
\end{center}
\vskip .15in
\centerline{\bf ABSTRACT}
\begin{quotation}
{\small We present a method for classifying conformal field theories based
on Coulomb gases (bosonic free-field construction). Given a particular
geometric configuration of the screening charges, we give necessary
conditions for the existence of degenerate representations and for the
closure of the vertex-operator algebra. The resulting classification
contains, but is more general than, the standard one based on classical Lie
algebras.  We then apply the method to the Coulomb gas theory for the
two-flavoured loop model of Jacobsen and Kondev.  The purpose of the study
is to clarify the relation between Coulomb gas models and conformal field
theories with extended symmetries.  }
\end{quotation}
\end{titlepage}

\newpage

\section{Introduction}

Conformal Field Theories (CFT) based on Coulomb gases arise in a variety of
two-dimensional problems in statistical physics \cite{ref1}. Furthermore, a
great number of CFTs are known to possess a Coulomb gas formulation. Early
examples, such as the Potts model or the critical and tricritical points in
the O($n$) model, were based on a single scalar field, and the physical
operators could be interpreted as particles carrying scalar quantised
electric and magnetic charges.

More recently, multicomponent Coulomb gases employing several bosonic free
fields have appeared in the study of critical phases in the so-called
fully-packed loop models \cite{ref2}. A first step in their resolution
consists in bijectively mapping configurations of oriented loops to those
of a discretised surface. The basic idea is here to interpret the loops as
contour lines of the surface height, but due to the fully-packing
constraint it turns out that the height variables in general have to be
vector valued.  Based on symmetry and entropic considerations, an effective
action of the Liouville type can then be written down for the continuum
limit of this interfacial representation. However, this action contains a
certain number of elastic constants whose exact values cannot be inferred
directly from the discrete model.

Important technical progress was achieved with the use of the {\em loop
ansatz}. It states that the most relevant vertex operators in a given model
have to be exactly marginal and taken into the action as screenings, thus
allowing all the elastic constants to be fixed \cite{ref3}. This situation
is truly remarkable: The discrete model precludes an a priori knowledge of
the parameters defining its continuum limit, but it nevertheless fixes the
geometry of the screening charges and thus permits an exact a posteriori
determination of the very same parameters. As the end result one obtains a
CFT in the form of a Coulomb gas (in general multicomponent, the number of
components being the dimensionality of the height space) with a given
background charge and screening operators. Physical operators are
represented by vertex operators
\beq
  V_{\vec{\beta}}(z,\bar{z})= :e^{i\vec{\beta}\vec{\varphi}(z,\bar{z})}:,
\eeq
with $\vec{\beta}$ taking values on a particular lattice, specific to the
given model. The vector $\vec{\varphi}(z,\bar{z})$ contains a set of
bosonic free fields $\vec{\varphi}(z,\bar{z})=\{\varphi_{1}(z,\bar{z}),
\varphi_{2}(z,\bar{z}),\ldots,\varphi_{D}(z,\bar{z})\}$, which can be
interpreted as the continuum limit of the components of the discrete
interfacial height.

The question which appears naturally is whether there is a chiral algebra
hidden behind this conformal theory, defined as a multicomponent Coulomb
gas, and what it might be. In particular, given the fact that fully-packed
loop models may possess a central charge $c>1$ \cite{ref2}, one may wonder
if the chiral algebra could be bigger than just the Virasoro one, ensured
by the conformal invariance of the model. Put differently: Might a
particular model, represented in the continuum limit as a Coulomb gas, have
additional, or extended symmetries? By this we mean not just global
symmetries, which are usually explicit, but symmetries which are extended
on the level of the chiral algebra, i.e.~extended infinitely. In case of a
positive answer this would imply the existence of extra chiral operators,
like $W(z)$ operators, which form an extended chiral algebra together with
the stress-energy tensor $T(z)$.

Presence of such symmetries in the Coulomb gas would not automatically
ensure that they persist in the original model defined on the lattice, and
that the model would be integrable. Still, in case of a positive answer for
the continuum theory (existence of extended symmetries) but negative for
the lattice model, one could try to find an integrable version of it on the
lattice.

On the other hand, in case of a negative answer in the continuum (no extra
chiral operators in the multicomponent Coulomb gas model, in addition to
$T(z)$) the prediction for the lattice model is likely to be definitive:
Extended symmetries will not come about on the lattice, and the model will
not be integrable. It is in this perspective, looking for the existence of
extra symmetries or their absence, that we shall try to clarify the
relation between Coulomb gas models and CFTs which are based on extended
chiral algebras.

The general method is presented in Section~\ref{sec2}. We here study the
necessary conditions for the existence of degenerate representations and
for the closure of the vertex-operator algebra. These conditions are of
course met by all W-type extended CFTs which are based on simply laced
Lie algebras, but in fact they are less restrictive. In particular they are
fulfilled by the two-flavour fully-packed loop model of Jacobsen and Kondev
\cite{ref2}, whose continuum limit is reminiscent of, but not identical to,
the classical WA${}_3$ theory. To assess whether such theories actually
contain extra chiral operators we work on a case-to-case basis, computing
the exact form of such operators or proving that they do not exist. In
Section 3 this method is applied to a variety of two- and three-component
Coulomb gases, in particular to those which arise in the solution of loop
models. We finally present our conclusions in Section 4.

\section{General method}
\label{sec2}

\subsection{Background charge and screenings}

The primary information defining a Coulomb gas model can be summarised as
follows:
\begin{enumerate}
 \item Dimensionality of the Coulomb gas (number of free-field components),
 $D$.
 \item The background charge, $\vec{\alpha}_{0}$.
 \item A set of $D$ screening operators,
\beq
       \{:\e^{i\vec{\alpha}_{a}\vec{\varphi}(z,\bar{z})}: | \ 
       a=1,2,\ldots,D \}. 
\label{screening}
\eeq
\end{enumerate}
We shall assume that, for a generic model, the number of screenings equals
the dimensionality of the Coulomb gas. Namely, if the number of screenings
were less than $D$ then a subset of free fields would decouple, and the
model could be represented as a direct product of a submodel with the
background charge and screenings and a submodel of actual free fields, no
background charge, and no screenings. If, alternatively, due to a special
symmetry of the lattice to which the screenings belong, the number of
screenings exceeds $D$, a subset of exactly $D$ screenings ought to be
sufficient to define correlation functions of physical operators. Different
choices of a subset of $D$ screenings should give equivalent results. At
least this should be the case for a Coulomb gas with an underlying chiral
algebra, which classifies all the operators in the model, and with respect
to which the physical operators are primaries. This last point will be made
more precise in the following. For the moment we do not have any reasons
for assuming such special symmetries of the lattice to which the screenings
should belong.

\begin{figure}
\begin{center}
 \leavevmode
 \epsfysize=140pt{\epsffile{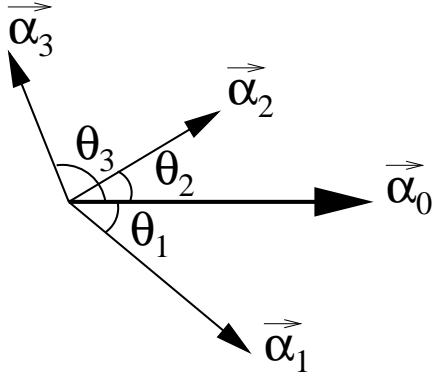}}
 \end{center}
 \protect\caption[3]{\label{fig1}Background charge vector $\vec{\alpha}_{0}$
 and the vectors $\{ \vec{\alpha}_a \}$ defining the screening operators
 $V_{\vec{\alpha}_a}(z,\bar{z})=
 \ : \e^{i\vec{\alpha}_{a}\vec{\varphi}(z,\bar{z})} :$
 for $a=1,2,\ldots,D$. On the figure, $D=3$.
 $\{\Theta_a\}$ are the angles which $\{ \vec{\alpha}_a \}$ make with
 the background charge $\vec{\alpha}_{0}$.}
\end{figure}

The primary information on a Coulomb gas, as stated above, is sketched in
Fig.~\ref{fig1}. For various lattice models, like Potts or O($n$) models
\cite{ref1}, or more general loop models \cite{ref2}, the geometry of
$(\{\vec{\alpha}_{a}\},\vec{\alpha}_{0})$ is known explicitly. But we shall
here derive general constraints on the possible geometries, by using
consistency conditions of the corresponding CFT.

We take $\vec{\alpha}_{0}$ as given. The stress-energy tensor of $D$ free
fields $\{\varphi_{a}(z,\bar{z})\}$ will be taken in the form
\beq
\label{stress}
 T(z)=-\frac{1}{4}:\partial\vec{\varphi}(z)\partial\vec{\varphi}(z):+
 i\vec{\alpha}_{0} \partial^2 \vec{\varphi}(z),
\eeq
with the two-point correlation functions of the fields
$\{\varphi_{a}(z,\bar{z})\}$ normalised as
\beq
 \langle \varphi_{a}(z,\bar{z})\varphi_{b}(z',\bar{z}') \rangle =
 2 \delta_{a,b} \log\frac{1}{|z-z'|^{2}}.
 \label{twopoint}
\eeq

With this normalisation, the conformal dimension
(with respect to the stress-energy tensor (\ref{stress})) of a vertex operator
\beq
\label{vertex}
 V_{\vec{\alpha}}(z,\bar{z})= \ :\e^{i\vec{\alpha}\vec{\varphi}(z,\bar{z})}:
\eeq
will be equal to
\beq
 \Delta_{\vec{\alpha}}=\vec{\alpha}^{2}-2\vec{\alpha}\vec{\alpha}_{0}.
 \label{dimension}
\eeq

The first condition on the screening operators (\ref{screening}) is that
they have to be marginal, $\Delta_{\vec{\alpha}_{a}}=1$ for $a=1,2,\ldots,D$.
This condition ensures that contour integrals of the screenings
\beq
 Q_{a}=\oint_{C}dzV_{\vec{\alpha}_{a}}(z),\quad a=1,2,...,D
 \label{commute}
\eeq
commute with the Virasoro algebra generated by $T(z)$. The operator
$V_{\vec{\alpha}_{a}}(z)$ in (\ref{commute}) is assumed to be the
holomorphic part of the screening operator
$V_{\vec{\alpha}_{a}}(z,\bar{z})$ in (\ref{screening}), in the sense of
holomorphic-antiholomorphic factorisation of correlation functions of
vertex operators, or, more generally, in the sense of
holomorphic-antiholomorphic factorisation of two-dimensional integrals of
correlation functions of vertex operators with respect to the contour
integrals \cite{ref4,ref5}. 

\begin{figure}
\begin{center}
 \leavevmode
 \epsfysize=140pt{\epsffile{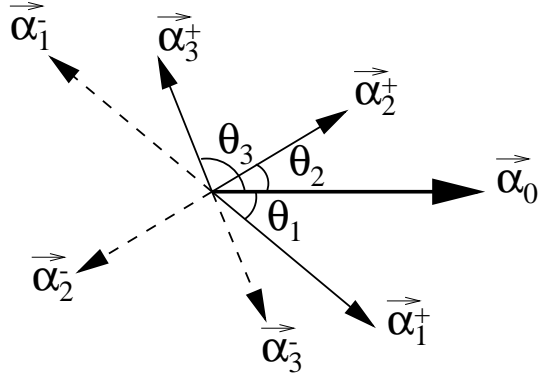}}
 \end{center}
 \protect\caption[3]{\label{fig2}The complete set of screenings
 $\{ \vec{\alpha}_a^+, \vec{\alpha}_a^- \}$, and
 the background charge $\vec{\alpha}_{0}$.}
\end{figure}

To make the screenings marginal, the vectors $\{\vec{\alpha}_{a}\}$ in
(\ref{screening}) have to satisfy the condition
\beq
  \Delta_{\vec{\alpha}_{a}}=\vec{\alpha}^{2}_{a}-2\vec{\alpha}_{a}
  \vec{\alpha}_{0}=1,
  \label{marginal}
\eeq
or
\beq
 \alpha^{2}_{a}-2\alpha_{a}\alpha_{0}\cos\Theta_{a}=1,
 \label{length1}
\eeq
\beq
 \alpha^{\pm}_{a}=\alpha_{0}\cos\Theta_{a}\pm\sqrt{1+
 \alpha^{2}_{0}\cos^{2}\Theta_{a}}.
 \label{length2}
\eeq
Here $\{\Theta_{a}\}$ are the angles in Fig.~\ref{fig1}, and
$\alpha^{\pm}_{a}$ are the ``lengths'' of the
vectors $\vec{\alpha}_{a}$ (one positive, $\alpha^{+}_{a}$, and one negative,
$\alpha^{-}_{a}$), which satisfy the condition (\ref{marginal}).

In Fig.~\ref{fig1} we have indicated a particular geometry of the vectors
$\vec{\alpha}_{0}$ and $\{\vec{\alpha}_{a}\}$. This choice of geometry is
for the moment arbitrary. The only constraint, once the directions of
$\vec{\alpha}_{0}$ and $\{\vec{\alpha}_{a}\}$ have been chosen,
is on the lengths of the screenings, Eq.~(\ref{length2}). There are two
solutions, ``$+$'' and ``$-$'', for each direction, and we shall see in the
following that the consistence of the corresponding CFT requires the use of
them both.
Since the ``lengths'' $\{\alpha^{-}_{a}\}$ are negative, the screening vectors
$\{\vec{\alpha}^{-}_{a}\}$ are oriented in the opposite direction with respect
to $\{\vec{\alpha}^{+}_{a}\}$. The set of screenings has thus been doubled,
$D$ screenings $\{\vec{\alpha}^{+}_{a}\}$ and $D$ screenings
$\{\vec{\alpha}^{-}_{a}\}$, as indicated in Fig.~\ref{fig2}.
We also note the relations between $\alpha^{+}_{a}$ and $\alpha^{-}_{a}$
\bea
 \alpha^{+}_{a}+\alpha^{-}_{a}&=&2\alpha_{0}\cos\Theta_{a}
 \label{trace} \\
 \alpha^{+}_{a}\alpha^{-}_{a}&=&-1,
 \label{determinant}
\eea
which follow from Eq.~(\ref{length1}).
These relations will be used in the following.

\subsection{Necessary condition for degenerate representations. Kac formula}
\label{sec:kacf}
We shall be looking for a Coulomb gas theory based on a chiral algebra.
The number of chiral operators forming this algebra has to be equal to the
number of free fields, $D$. In addition to $T(z)$, this requires $D-1$
extra chiral operators to control all the degrees of freedom of the theory,
i.e.~to control $D$ free fields.

We shall assume that these extra chiral operators belong to the module of
the identity operator, just like $T(z)$, i.e.~that they are made as linear
combinations of products of derivatives of free fields. For a Coulomb gas
theory based on $D$ bosonic fields this last assumption appears to us to be
natural. This is essentially because in this case the chiral operators will
have zero Coulomb charge; their correlation functions will therefore not
require the integrated screening operators, they will be ``simple''.

It should be noted that this argument is valid only in the case of
$\vec{\alpha}_{0} \neq 0$, where, in general, correlation functions of
generic vertex operators {\em do} require screenings. Generic correlation
functions are therefore not simple, but those made exclusively of chiral
symmetry operators are. This argument implicitly supposes that an
acceptable Coulomb gas theory will allow for a deformation of the trivial
free-field point $\vec{\alpha}_{0} = 0$, which has $c=D$.

Assuming therefore that the extra chiral operators belong to the module of
the identity, the possible chiral algebras for Coulomb gas theories should
be made of operators with integer conformal dimensions, $\Delta \in \N$.
Furthermore, since a dimension-one current is known to generate a
continuous rather than a discrete symmetry, we require $\Delta\geq 2$. This
means that the chiral algebras should be of the W type. If this algebra
exists, the screening operators (integrated along the contours) have to
commute with it, in order to respect the extended symmetry of the given
theory.

For chiral operators in the form of linear combinations of products of
derivatives of free fields, the operators which are primary with respect
to them are the vertex operators
\beq
 V_{\vec{\beta}}(z,\bar{z})=\ : \e^{i\vec{\beta}\vec{\varphi}(z,\bar{z})} :,
 \label{vertex1}
\eeq
where $\vec{\beta}$ is for the moment arbitrary. In fact, an operator
$\Phi(z,\bar{z})$ being primary with respect to a particular chiral
operator $W(z)$ means that the operator product expansion (OPE) of $W(z)$
with $\Phi(z,\bar{z})$ takes the form
\beq
 W(z)\Phi(z',\bar{z}')=\frac{A}{(z-z')^{\Delta_{W}}}\Phi(z',\bar{z}')+\ldots,
 \label{OPE}
\eeq
where $A$ is a (structure) constant. Loosely speaking, the most singular term
in the expansion in powers of $(z-z')$ produces again the operator
$\Phi(z,\bar{z})$. Evidently, when $W(z)$ is a linear combination of products
of derivatives of free fields, each term involving a total of $\Delta_{W}$
derivatives, and when
$\Phi(z,\bar{z})=V_{\vec{\beta}}(z,\bar{z})$ is an exponential of free fields,
cf.~Eq.(\ref{vertex1}), the expansion $W(z)V_{\vec{\beta}}(z',\bar{z}')$ will
be precisely of the form (\ref{OPE}).

\begin{figure}
\begin{center}
 \leavevmode
 \epsfysize=80pt{\epsffile{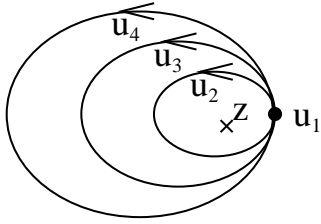}}
 \end{center}
 \protect\caption[3]{\label{fig3}Having integrated first all the screenings
 except for one type (whose choice is not important), one finally
 integrates the last screening, at the point $u_1$, around $z$.}
\end{figure}

Until now $\vec{\beta}$ has been an arbitrary vector. For special values of
$\vec{\beta}$ it may however happen that primary operators appear in the
module of $V_{\vec{\beta}}(z,\bar{z})$. They are called singular states, and
their presence means degeneracy of the module. Since the integrated screenings
commute with the operators of the chiral algebra, the singular state in the
module of $V_{\vec{\beta}} (z,\bar{z})$ may be obtained by mapping, with the
use of screenings, another vertex operator $V_{\vec{\beta}'}(z,\bar{z})$
into the module of $V_{\vec{\beta}}(z,\bar{z})$.
This is brought about in the following way:
\beq
 S(V_{\vec{\beta}}(z,\bar{z}))=
 \prod^{D}_{a=1} \left( \oint_{C_{a}}du_{a}V_{\vec{\alpha}^{+}_{a}}
 (u_{a}) \right)^{n_{a}}V_{\vec{\beta}'}(z,\bar{z}),
 \label{singular}
\eeq
where $n_a \in \N_0$ for $a=1,2,\ldots,D$. Here $S(V_{\vec{\beta}})$ is
a singular state in the module of $V_{\vec{\beta}}$,
and $\vec{\beta}'$ should be related to $\vec{\beta}$ by
\beq
 \vec{\beta}'=\vec{\beta}-\sum_{a}n_{a}\vec{\alpha}^{+}_{a}.
 \label{singular1}
\eeq
The configuration of contours $\{C_{a}\}$ in Eq.~(\ref{singular}) is shown
in Fig.~\ref{fig3}. The set of contours in the figure is only schematic,
since there are in fact $n_a$ distinct contours corresponding to the screenings
of type $a=1,2,\ldots,D$.

This prescription for realising closed contour integrations of screenings
around a fixed vertex operator, on which they act to induce the mapping
(\ref{singular}), was first used in Ref.~\cite{ref6}.

More precisely, the mapping of Eq.~(\ref{singular}) is defined,
and the singular state $S(V_{\vec{\beta}})$ exists, only if
\beq
 \Delta_{\vec{\beta}'}-\Delta_{\vec{\beta}}=N,
 \label{condition}
\eeq
where $N$ is a positive integer.  In fact, as the screenings commute with
$T(z)$, the conformal dimension of $S(V_{\vec{\beta}})$ is equal to that of
$V_{\vec{\beta}'}$. Being a descendent state, the conformal dimension of
$S(V_{\vec{\beta}})$ should differ from $\Delta_{\vec{\beta}}$ by an
integer.  One thus obtains the condition (\ref{condition}).

Another way to get (\ref{condition}) is to require that the monodromy for
the analytic continuation of the expression on the right-hand side of
(\ref{singular}) with respect to the common variable $u_{1}$
(cf.~Fig.~\ref{fig3}) be trivial, so that the contour of integration over
$u_{1}$ around $z$ closes. Otherwise the integrated screenings would not
actually commute with $T(z)$ and with other operators of the chiral
algebra.  It can be checked that this trivial monodromy requirement leads
equivalently to the condition (\ref{condition}) \cite{ref6}.

This condition, with $\vec{\beta}'$ defined by (\ref{singular1}) and
$\Delta_{\vec{\beta}}, \Delta_{\vec {\beta}'}$ given by (\ref{dimension}),
fixes special values of $\vec{\beta}$ for which the module of
$V_{\vec{\beta}}$ contains singular states, i.e.~is degenerate.

Note that we have only applied the ``$+$'' screenings in
Eqs.~(\ref{singular})--(\ref{singular1}). The reason is that once the
solution for $\vec{\beta}$ is found, it will automatically satisfy the
corresponding condition with negative screenings.  For the analysis of
singular states the ``$+$'' and ``$-$'' screenings are therefore
complimentary. A similar remark holds true for a mixed mapping, i.e.~one
involving both ``$+$'' and ``$-$'' screenings: The corresponding conditions
do not lead to new solutions.  These statements should become more clear
from the analysis that will follow.

We remark that this way of defining the degenerate representations, using
the mapping by screenings between vertex operators and vertex operator
modules, has probably been used first in Ref.~\cite{ref7} to reproduce the
Kac formula for conformal dimensions for the degenerate representations of
the Virasoro algebra. This method has also been used in \cite{ref8,ref9} to
define the degenerate representations of W-algebra CFTs based on
classical Lie algebras.

Until now we have assumed a set of screenings, as depicted in
Figs.~\ref{fig1}--\ref{fig2}, with no restrictions yet, except for those on
the lengths of screenings, cf.~Eqs.~(\ref{marginal})--(\ref{length2}). As
we are looking for Coulomb gas theories which should eventually be endowed
with extended chiral algebras, we shall next identify the vertex operators
$\{V_{\vec{\beta}}(z,\bar{z})\}$ which induce the degenerate
representations of that chiral algebra.

Our approach can be outlined as follows: Without knowing yet the chiral
algebra, for a given set of screening operators we shall define, by using
the mappings described above, the ``would be degenerate
representations''. The screening operators are assumed to commute with the
(presently unknown) chiral algebra. We shall define, for given screenings
and background charge, the discrete set of values of $\vec{\beta}$ for
which the operators $V_{\vec{\beta}}(z,\bar{z})$ would induce degenerate
representations, if the chiral algebra really existed. In this way we shall
arrive at a necessary condition for the existence of degenerate
representations. The condition will also become sufficient when the
corresponding extended chiral algebra is found.

The advantage of this approach will be that, having defined such special
values of $\vec{\beta}$ corresponding to given screenings, it will allow us
to derive next a set of constraints ``acting backwards'', i.e.~that limits
the allowed configurations of screenings and the background charge.

{}From the vast literature on W-algebras it is well-known that the
proper way to look for singular states in the module of
$V_{\vec{\beta}}(z, \bar{z})$ is to apply mappings of the type
(\ref{singular}) by shifting $\vec{\beta}'$ away from $\vec{\beta}$ only
along the ``principal'' directions defined by the individual screenings,
and not as has been sketched schematically in
Eqs.~(\ref{singular})--(\ref{singular1}), where several different
screenings are involved simultaneously.

\begin{figure}
\vskip 3cm
\hskip 6cm
 \epsfysize=140pt{\epsffile{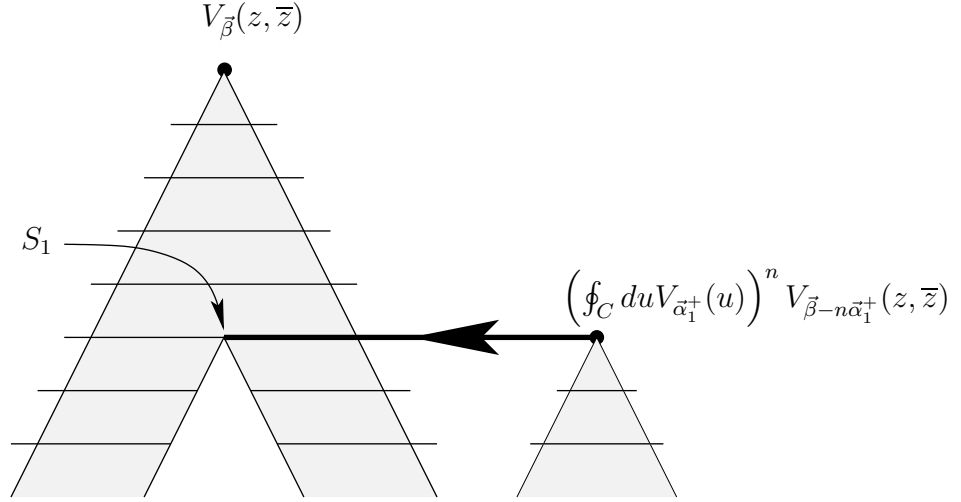}}
\vskip -8.0cm \hskip 6.2cm $V_{\vec{\beta}}(z,\overline{z})$
\vskip 2.2cm  \hskip 3.8cm $S_1$
\vskip 0.cm \hskip 11cm $\left(\oint_C du V_{\vec{\alpha}_1^+}(u)\right)^n
V_{\vec{\beta} -n\vec{\alpha}_1^+}(z,\overline{z})$
\vskip 3.5cm
\protect\caption[3]{\label{fig4}Mapping of the module of
$V_{\tilde{\vec{\beta}}} \equiv V_{\vec{\beta}-n\vec{\alpha}_1^+}$
into the module of $V_{\vec{\beta}}$, in the direction of the first
screening $\vec{\alpha}_1^+$.}
\end{figure}

For the direction associated with the first screening,
the mapping takes the form
\bea
 S_{1}(V_{\vec{\beta}}(z,\bar{z}))&=&
 \left( \oint_{C}duV_{\vec{\alpha}^{+}_{1}}(u) \right)^{n}V_{\vec{\beta}-
 n\vec{\alpha}^{+}_{1}}(z,\bar{z})\nn\\
 &=& \prod^{n}_{i=1}\oint du_{i}V_{\vec{\alpha}_{1}^{+}}
 (u_{i})V_{\vec{\beta}-n\vec{\alpha}^{+}_{1}}(z,\bar{z}),
\eea
cf.~Figs.~\ref{fig3}--\ref{fig4}. The condition (\ref{condition}) becomes
\beq
 \Delta_{\vec{\beta}-n\vec{\alpha}^+_{1}}-\Delta_{\vec{\beta}}=N,
 \label{condbeta}
\eeq
where $N$ should be a positive integer. Substituting the formula
(\ref{dimension}) for $\Delta_{\vec{\beta}-n\vec{\alpha}^+_{1}}$
and $\Delta_{\vec{\beta}}$, one finds
\beq
 (\vec{\beta}-n\vec{\alpha}^{+}_{1})^{2}-2\vec{\alpha}_{0}(\vec{\beta}-
 n\vec{\alpha}_{1}^{+})-
 \vec{\beta}^{2}+2\vec{\alpha}_{0}\vec{\beta}=N,
\eeq
which gives
\beq
 -2n(\vec{\beta}-\vec{\alpha}_{0})\vec{\alpha}_{1}^{+}+n^{2}
 (\alpha_{1}^{+})^{2}=N.
 \label{eqbeta}
\eeq
The general solution of this equation for the allowed values of $\vec{\beta}$
takes the form
\beq
 2(\vec{\beta}-\vec{\alpha}_{0})=(n\alpha^{+}_{1}+n'\alpha^{-}_{1})
 \vec{\omega}_{1},
 \label{solbeta}
\eeq
where the vector $\vec{\omega}_{1}$ should verify the equation
$\vec{\omega}_{1} \vec{e}_{1}=1$
with
$\vec{e}_{1}=\vec{\alpha}^{+}_{1}/\alpha^{+}_{1}$,
and $n'$ in (\ref{solbeta}) is an another positive integer.
By substituting (\ref{solbeta}) into (\ref{eqbeta}) one finds
$N=nn'$
so that, for $\vec{\beta}$ of the form (\ref{solbeta}),
the constraint (\ref{condbeta}) is indeed verified.

By similarly requiring that singular states be produced for each of the $D$
screening directions, defined by the vectors $\{\vec{\alpha}^{+}_{a}\}$, one
finds that $\vec{\beta}$ has to satisfy
\beq
 2(\vec{\beta}-\vec{\alpha}_{0})=
 \sum^{D}_{a=1}(n_{a}\alpha_{a}^{+}+n_{a}'\alpha^{-}_{a})
 \vec{\omega}_{a}.
 \label{solbeta1}
\eeq
Here the $D$ vectors $\{\vec{\omega}_{a}\}$ are defined by
\beq
 \vec{\omega}_{a}\vec{e}_{b}=\delta_{a,b},
 \label{ortho}
\eeq
and $\{\vec{e}_{a}\}$ are the unit vectors which define the directions
of the screenings:
\beq
 \vec{\alpha}_{a}^{\pm}=\alpha^{\pm}_{a}\vec{e}_{a} \mbox{ for } a=1,2,...,D.
 \label{unit}
\eeq

Eq.~(\ref{solbeta1}) generalises (\ref{solbeta}). Alternatively it can be
presented as
\beq
 \vec{\beta}=\vec{\beta}_{(n'_{1},n_{1})(n'_{2},n_{2})\cdots(n'_{D},n_{D})}=
 \sum^{D}_{a=1}
 \left( \frac{n_{a}}{2}\alpha^{+}_{a}+\frac{n'_{a}}{2}\alpha_{a}^{-} \right)
 \vec{\omega}_{a}+\vec{\alpha}_{0}.
 \label{alternative}
\eeq
This expression can be simplified by developing $\vec{\alpha}_{0}$
in terms of the vectors $\{\vec{\omega}_{a}/2\}$,
\beq
 \vec{\alpha}_{0}=\sum_{a}X_{a}\frac{\vec{\omega}_{a}}{2}.
 \label{develop}
\eeq
The expansion coefficients $\{X_{a}\}$ are then found by using the
orthogonality property (\ref{ortho}):
\beq
 X_{b}=2 \vec{\alpha}_{0}\vec{e}_{b} =
 2\alpha_{0}\cos \Theta_{b}=\alpha^{+}_{b}+\alpha^{-}_{b},
 \label{coefs}
\eeq
where in the last step we have used the relations (\ref{trace}) between
$\alpha^{+}_{a}$ and $\alpha^{-}_{a}$.
The angles $\{\Theta_{a}\}$ have been defined in Figs.~\ref{fig1}--\ref{fig2}.
Substituting (\ref{coefs}) into (\ref{develop}), we arrive at
\beq
 \vec{\alpha}_{0}=
 \sum_{a}\frac{\alpha^{+}_{a}+\alpha^{-}_{a}}{2}\vec{\omega}_{a},
 \label{background}
\eeq
which can finally be inserted in Eq.~(\ref{alternative}) to yield
the following set of values of $\vec{\beta}$:
\beq
 \vec{\beta}_{(n'_{1},n_1)(n'_{2},n_{2})\cdots(n'_{D},n_{D})}=
 \sum^{D}_{a=1} \left( \frac{1+n_{a}}{2}
 \alpha^{+}_{a}+\frac{1+n'_{a}}{2}\alpha^{-}_{a} \right) \vec{\omega}_{a}.
 \label{beta}
\eeq
For the ``conjugate'' values of $\vec{\beta}$,
viz.~$\tilde{\vec{\beta}}=2\vec{\alpha}_{0}-\vec{\beta}$, one finds:
\beq
 \tilde{\vec{\beta}}_{(n'_{1},n_{1})(n'_{2},n_{2})\cdots(n'_{D},n_{D})}=
 \sum^{D}_{a=1}
 \left(\frac{1-n_{a}}{2}\alpha^{+}_{a}+\frac{1-n'_{a}}{2}\alpha^{-}_{a}\right)
 \vec{\omega}_{a}.
 \label{beta1}
\eeq

\begin{figure}
\vskip 3cm
\hskip 6cm
 \epsfysize=140pt{\epsffile{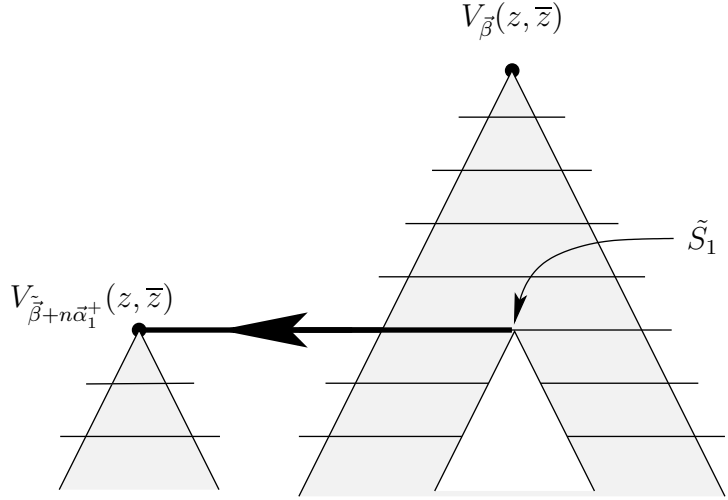}}
\vskip -8.0cm \hskip 9cm $V_{\vec{\beta}}(z,\overline{z})$
\vskip 2.2cm  \hskip 12.cm $\tilde{S_1}$
\vskip 0.cm \hskip 3cm $V_{\tilde{\vec{\beta}}
+n\vec{\alpha}_1^+}(z,\overline{z})$ 
\vskip 3.5cm
 \protect\caption[3]{\label{fig5}The mapping is of the form
 $\left(\oint_C {\rm d}u \, V_{\vec{\alpha}_1^+}(u) \right)^n \tilde{S}_1 =
 V_{\tilde{\vec{\beta}}+n \vec{\alpha}_1^+}$.}
\end{figure}

The manifestation of singular states in the module of the vertex operator
$V_{\tilde{\vec{\beta}}}(z,\bar{z})$ is so to say dual to that of
$V_{\vec{\beta}}(z,\bar{z})$. If the module of $V_{\tilde{\vec{\beta}}}$ is
degenerate, it contains particular primary states which produce, as images
under appropriate mappings, the vertex operators $V_{\vec{\beta}'}$ several
screenings away, as indicated in Fig.~\ref{fig5}. More details on these
properties can be found in the paper \cite{ref6}. Physically, in theories
with a chiral algebra, the set of dual operators
$\{V_{\vec{\beta}},V_{\tilde{\vec{\beta}}}\}$ can either represent the same
(singlet) operator (such as the energy operator) or a doublet of conjugate
operators (an example being the $\{\sigma, \sigma^{\dagger}\}$ spin
operators in the three-state Potts model).

For the conformal dimensions of the vertex operators
$V_{\vec{\beta}}(z,\bar{z})$, with the above values for $\vec{\beta}$,
one obtains
\bea
 \Delta_{\vec{\beta}_{(\cdots)}}&=&
 \Delta_{2\vec{\alpha}_{0}-\vec{\beta}_{(\cdots)}}\equiv
 \Delta_{(n'_{1},n_1)(n'_{2},n_{2})\cdots(n'_{D},n_{D})} \\
 &=& \sum^{D}_{a=1} \left( u^{(a)}_{n'_{a},n_{a}} \right)^{2}
 (\vec{\omega}_{a})^{2} + 2\sum^{D}_{a<b}u^{(a)}_{n'_{a},n_{a}}
 u^{(b)}_{n'_{b},n_{b}} \, \vec{\omega}_{a} \vec{\omega}_{b} -
 \sum^{D}_{a=1}\sum^{D}_{b=1}
 (\alpha^{+}_{a}+\alpha^{-}_{a})u^{(a)}_{n'_{a},n_{a}} \,
 \vec{\omega}_{a} \vec{\omega}_{b}, \nn
 \label{Kac}
\eea
where we have defined
\beq
 u^{(a)}_{n'_{a},n_{a}}=\frac{1-n_{a}}{2} \, \alpha^{+}_{a} +
 \frac{1-n'_{a}}{2} \, \alpha^{-}_{a}.
\eeq

The procedure that we followed above to define
$\beta_{(n'_{1}n_{1})(n'_{2}n_{2})\cdots(n'_{D}n_{D})}$ in
Eqs.(\ref{beta})--(\ref{beta1}), is perfectly standard. Initially it has been
used in Refs.~\cite{ref7,ref8,ref9}.

Note that if the chiral algebra consisting of $D$ chiral operators
commuting with the screenings is found, Eq.~(\ref{Kac}) together with
Eqs.(\ref{beta})--(\ref{beta1}) would become the Kac formula for the
degenerate representations of the Coulomb gas theory at hand.  But in our
approach it should rather be interpreted as a {\em necessary condition} for
the existence of degenerate representations, for any given configuration of
the screenings. The existence of the extended chiral algebra is not yet
guaranteed.

It should be noticed that the degeneracy of the module of $V_{\vec{\beta}}$
in all $D$ screening directions, with $\vec{\beta}$ taking its values in
the discrete set (\ref{beta}), must be required in order for the operator
algebra of the operators $\{V_{\vec{\beta}}(z,\bar{z})\}$ to be well
defined. We shall consider the properties of this vertex-operator algebra
shortly.

Another remark is on a detail which was actually implicit in the above
arguments: For a theory with an extended chiral algebra to be acceptable we
want it to contain one free parameter. Put differently, we want the theory
to exist for general values of $\alpha_{0}$, or, alternatively, of the
central charge $c$. In the class of CFTs in which we are interested, $c$
must remain a free parameter until we eventually impose the much stronger
constraint of unitarity, in which case it will be forced to take its values
in a discrete (but still denumerably infinite) set. In general we expect
the structure of a CFT to be very rigid, in the sense that once we impose
additional constraints that limit the generality of the theory, we will
immediately either generate inconsistencies or, in the best case, fix $c$
to take a finite number of values. We can illustrate this remark through
the example of the two-flavoured loop model of Jacobsen and Kondev
\cite{ref2}. In the special cases where the fugacities of either loop
flavour take the value 2 (resp.~1), the model is known to be equivalent to
the su(4)${}_{k=1}$ Wess-Zumino-Witten model \cite{Kondev-su4} (resp.~to
the equally-weighted six-vertex model), both of which are bona fide CFTs
with fixed $c=3$ (resp.~$c=1$). But it is not because of these two special
cases that we shall accept the general model (i.e.~with generic values of
the loop fugacities) as a theory with an extended chiral algebra.

Returning to the general study, one could define the central charge from the
two-point function of $T(z)$
\beq
 \langle T(z)T(z') \rangle =\frac{c/2}{(z-z')^{4}},
 \label{central}
\eeq
where $T(z)$ is given by (\ref{stress}), and the correlation functions of the
fields $\{\varphi_{a}(z,\bar{z})\}$ are normalised as in Eq.~(\ref{twopoint}). 
{}From (\ref{stress}), (\ref{twopoint}) and (\ref{central}) one then finds
\beq
 c=D-24 (\alpha_0)^2.
\eeq

The Coulomb gas theory itself is defined for general values of
$\alpha_{0}$.  Naturally, then, we are looking for conditions on the theory
under which it would contain an extended chiral algebra for general values
of $\alpha_{0}$, or, equivalently, of the central charge.

\subsection{Closure of the operator algebra of the operators
$\{V_{\vec{\beta}_{(\cdots)}}\}$}

Having defined the ``would be degenerate'' operators
$\{V_{\vec{\beta}_{(...)}}\} \equiv
 \{V_{(n'_{1},n_{1})(n'_{2},n_{2})\cdots(n'_{D},n_{D})}\}$,
with $\{\vec{\beta}_{(\cdots)}\}$ defined by Eqs.~(\ref{beta})--(\ref{beta1}),
we shall impose next the condition that their operator algebra closes.
In a Coulomb gas theory with background charge and screenings
the OPE for a product of two vertex operators
$V_{\vec{\beta}_{1}}(z,\bar{z})$ and $V_{\vec{\beta}_{2}}(z',\bar{z}')$
produces by conservation of electric charge the vertex operator
$V_{\vec{\beta}_{1}+\vec{\beta}_{2}}(z',\bar{z}')$,
but also the vertex operators where the charge
$\vec{\beta}_{1}+\vec {\beta}_{2}$ has been shifted by any number of
screening vectors. Thus, schematically,
\bea
 V_{\vec{\beta}_{1}}(z,\bar{z})V_{\vec{\beta}_{2}}(z',\bar{z}') &\sim&
 V_{\vec{\beta}_{1}+\vec{\beta}_{2}}(z',\bar{z}')+\mbox{descendents}\nn\\
 &+&V_{\vec{\beta}_{1}+\vec{\beta}_{2}+\vec{\alpha}_{1}^{+}}(z',\bar{z}')
 +\mbox{descendents}+ \cdots.
 \label{ope}
\eea
In general the OPE of $V_{\vec{\beta}_{1}}V_{\vec{\beta}_{2}}$ will produce
the entire set of operators
\beq
 \{V_{\vec{\beta}_{1}+\vec{\beta}_{2}+\sum_{a}(k_{a}\vec{\alpha}_{a}^{+}+
 l_{a}\vec{\alpha}_{a}^{-})}\}
 \label{opset}
\eeq
as well as their descendents.
If $\vec{\beta}_{1}$ and $\vec{\beta}_{2}$ belong to the Kac table,
i.e.~to the sets
$\{\vec{\beta}_{(n'_{1},n_{1})(n'_{2},n_{2})\cdots(n'_{D},n_{D})}\}$,
$\{\tilde{\vec{\beta}}_{(n'_{1},n_{1})(n'_{2},n_{2})...(n'_{D},n_{D})}\}$
given by Eqs.~(\ref{beta})--(\ref{beta1}), then in order that the operators
(\ref{opset}) also belong to the Kac table, the screening vectors
$\{\vec{\alpha}^{+}_{a}\}$ must decompose as a integer linear combination
of the basis vectors spanning (\ref{beta})--(\ref{beta1}),
i.e.~the set
$\{\alpha^{+}_{a}\frac{\vec{\omega}_{a}}{2}\}$.
It is sufficient to ensure this decomposition for the ``$+$'' screenings,
since the corresponding decomposition of the ``$-$'' screenings over the
basis $\{{\alpha}^{-}_{a}\frac{\vec{\omega}_{a}}{2}\}$ will then follow
automatically. Writing
\beq
 \vec{\alpha}^{+}_{a}=\sum_{b}A_{ab}\alpha^{+}_{b}\frac{\vec{\omega}_{b}}{2},
\eeq
the integer coefficients $A_{ab}$ can be computed by projecting onto the
normalised screening vectors, cf.~Eq.~(\ref{unit}):
\bea
 \vec{\alpha}^{+}_{a} \vec{e}_{c} &=&
 \sum_{b}A_{ab}\alpha^{+}_{b}\frac{1}{2}
 \vec{\omega}_{b} \vec{e}_{c} =
 \sum_{b}A_{ab}\alpha^{+}_{b}\frac{1}{2}
 \delta_{bc}=\frac{1}{2}A_{ac}\alpha^{+}_{c} \\
 A_{ac} &=& \frac{2(\vec{\alpha}^{+}_{a},\vec{e}_{c})}{\alpha^{+}_{c}}=
 \frac{2\vec{\alpha}^{+}_{a} \vec{\alpha}^{+}_{c}}{(\vec{\alpha}^{+}_{c})^{2}}.
 \label{Cartan}
\eea
We observe that when expressing the matrix of coefficients $A_{ab}$ in terms
of screening vectors, one recovers the same form as the Cartan matrix in the
theory of classical Lie algebras, expressed in terms of the simple root
vectors. The classification condition is the same: Its components have to be
integers.

It is well-known that this condition implies that the screening vectors
$\{\vec{\alpha}^{+}_{a}\}$ must belong to the root lattice of one of the
classical Lie algebras. This criterion appears here as a {\em necessary
condition} for the closure of the operator algebra of the vertex operators
$V_{(n'_{1},n_{1})(n'_{2},n_{2})...(n'_{D},n_{D})}(z,\bar{z})$.  Clearly,
since we are interested in the most general Coulomb gas theory, and not
just a closed sub-theory, we can further require the
$\{\vec{\alpha}^{+}_{a}\}$ to be a set of {\em basis vectors} of the root
lattice. However, there is for the moment no need that we should constrain
further to the {\em simple roots} of a classical Lie algebra, in which case
we would immediately limit ourselves to the standard classification of
W-type extended CFTs.

Several remarks are in order.

In the Coulomb gas theory there are additional constraints on the lengths
of the vectors $\{\vec{\alpha}^{+}_{a}\}$, in order to ensure the
marginality of the screening operators. These constraints are expressed by
Eq.~(\ref{length2}).

For the case of simply laced algebras, all the lengths of $\{\vec{\alpha}
^{+}_{a}\}$ have to be equal. In this case, to maintain the marginality
constraint (\ref{length2}) upon varying $\alpha_{0}$, all the angles
$\{\Theta_{a}\}$ have to be equal.

For non-simply laced algebras, e.g.~$B_{n}$, the ratio of the lengths of
long and short roots is $\sqrt{2}$. This is incompatible with
Eq.~(\ref{length2}), at least for small values of $\alpha_{0}$. We conclude
that for Coulomb gas models defined in terms of bosonic fields only, the
screening vectors $\{\vec{\alpha}^{+}_{a}\}$ have to correspond to the
basic vectors of the root lattice of simply laced algebras only: A${}_{n}$,
D${}_{n}$, E${}_{6}$, E${}_{7}$ or E${}_{8}$.  It is in fact well-known that the
free-field representation for W-theories based on non-simply laced
algebras uses fermionic fields, in addition to the bosonic ones \cite{ref9}.

We emphasise once again that we are analysing the generic case of a Coulomb
gas model of $D$ bosonic fields $\{\varphi_{a}(z,\bar{z})\}$, with a
non-degenerate set of $D$ screening vectors $\{\vec{\alpha}^{+}_{a}\}$
which couple to {\em all} the fields $\{\varphi_{a}(z,\bar{z})\}$. On the
contrary, if the set of the screening operators is reduced and some of the
fields $\{\varphi_{a}(z,\bar{z})\}$ are decoupled from screenings, our
arguments do not apply. For instance one could imagine having an extra
bosonic field which bosonises a couple of fermionic fields. Evidently,
there is a wide range of possibilities of constructing non-generic Coulomb
gas models.

As has been concluded above, in our Coulomb gas models all
$\{\Theta_{a}\}$ have to be equal, i.e. the vector $\vec{\alpha}_{0}$
has to be ``equidistant'' from all the screening vectors,
cf.~Fig.~\ref{fig1}. All the lengths of the screening vectors
$\{\vec{\alpha}^{+}_{a}\}$ have to be identical, as have those of
$\{\vec{\alpha}^{-}_{a}\}$. We shall henceforth denote them as
$\alpha_{+}$ and $\alpha_{-}$.
Eqs.~(\ref{length2})--(\ref{determinant}) now take the simpler form
\bea
 \alpha_{\pm}&=&\alpha_{0}\cos\Theta\pm\sqrt{\alpha^{2}_{0}\cos^{2}\Theta+1},\\
 \alpha_{+}+\alpha_{-} &=& 2\alpha_{0}\cos\Theta, \\
 \alpha_{+}\alpha_{-} &=& -1.
\eea
Similarly, Eq.~(\ref{background}) for the background charge vector becomes
\beq
 2\vec{\alpha}_{0}=\sum_{a}(\alpha^{+}_{a}+\alpha^{-}_{a})\vec{\omega}_{a}
 =(\alpha_{+}+\alpha_{-})\sum_{a}\vec{\omega}_{a},
 \label{alpha}
\eeq
and Eq.~(2.44) now reads
\beq
 \vec{\alpha}^{\pm}_{a}=\sum_{b}A_{ab}\alpha^{\pm}_{b}\frac{\vec{\omega}_{b}}
 {2}=\frac{\alpha_{\pm}}{2}\sum_{b}A_{ab}\vec{\omega}_{b}.
\eeq
{}From this latter equation one finds
\beq
 \vec{\omega}_{a}=
 \sum_{b}A^{-1}_{ab}\frac{2\vec{\alpha}^{\pm}_{b}}{\alpha_{\pm}}
 =\sum_{b}2A^{-1}_{ab}\vec{e}_{b},
 \label{omega}
\eeq
where $A^{-1}_{ab}$ is a matrix inverse of $A_{ab}$. Substituting
(\ref{omega}) into (\ref{alpha}) one obtains
\beq
 2\vec{\alpha}_{0}=
 (\alpha_{+}+\alpha_{-})\sum_{a}\sum_{b}2A^{-1}_{ab}\vec{e}_{b} \equiv
 (\alpha_{+}+\alpha_{-})\sum_{b}m_{b}\vec{e}_{b},
\label{alphae}
\eeq
with the notation
$m_{b}=\sum_{a}2A^{-1}_{ab}$.
We finally arrive at the following relation between the background charge
and the screenings:
\beq
 2\vec{\alpha}_{0}=\sum_{a}m_{a}(\vec{\alpha}^{+}_{a}+\vec{\alpha}^{-}_{a}).
 \label{Cartan1}
\eeq
Multiplying by $\vec{\alpha}_0$ we also deduce that 
\beq
{1\over \cos^2\theta} = \sum_a m_a \; .
\eeq

Another remark can be made concerning the degeneracy assumption of the
module of $V_{\vec{\beta}}(z,\bar{z})$, which has been used throughout the
analysis. Its justification is that an extended chiral algebra must
necessarily possess degenerate representations realised by the vertex
operators, which are primaries as has been argued in the beginning of this
Section.

That the operator algebra of $\{V_{\vec{\beta}}\}$, corresponding to such
degenerate representations, should close can be shown by using, for
instance, the differential equations for correlation functions.  These
equations could be derived if the singular states were realised explicitly
by the chiral algebra operators. In the analysis we have just used these
assumptions.

One particular feature in the construction of ``would be singular states''
by mappings produced by the integrated screenings is that we have demanded
the degeneracy in all $D$ directions, corresponding to the $D$ vectors
$\{\vec{\alpha}^{+}_{a}\}$. This condition is necessary to ensure that the
OPE of two operators $V_{\vec{\beta}_{1}}(z,\bar{z})$,
$V_{\vec{\beta}_{2}}(z',\bar{z}')$ 
produces only a finite number of primary operators out of an, in principle, 
infinite set of operators:
\beq
 \{V_{\vec{\beta}_{1}+\vec{\beta}_{2}+\sum_{a}(k_{a}\vec{\alpha}^{+}_{a}+l_{a}
 \vec{\alpha}^{-}_{a})}\},
 \label{operators}
\eeq
cf.~the arguments given above in connection with
Eqs.~(\ref{ope})--(\ref{opset}).

The support for this way of implementing the degeneracy assumption can be
found, on one hand, in the fact that a finite OPE (in a sense of a finite
number of primaries appearing in the OPE) is also a consequence of the
above-mentioned differential equations.
On the other hand, a subset of the primaries produced by the OPE
(\ref{ope}) by adding an appropriate amount of screenings will consist of
vertex operators $V_{\vec{\beta}'}(z,\bar{z})$ which are preimages of
singular states, in the sense of the mapping (\ref{singular}). This subset,
which constitutes the unphysical part of the Kac table, have to decouple from
the rest in correlation functions as well as in the operator algebra of
physical operators. In this way the infinite set of operators appearing in
(\ref{operators}) will be restricted to a finite number of physical
operators. The fact that we demand degeneracy in all $D$ directions is thus
tantamount to bordering the physical domain of the Kac table in all these
directions.
%

It is interesting to apply the above result to the
two-flavoured loop model introduced by Jacobsen and Kondev \cite{ref2}.
Its configurations are those of two colours of closed 
loops ($N_{\rm b}$ black loops and $N_{\rm g}$ grey ones) placed on the
edges of a square lattice, in such a way that every vertex of the lattice
touches exactly one loop of either colour. Introducing loop fugacities
$n_{\rm b}$ and $n_{\rm g}$ for the loop flavours, the partition function
of the discrete model reads
\beq
 Z = \sum_{\cal G} n_{\rm b}^{N_{\rm b}} n_{\rm g}^{N_{\rm g}},
\eeq
where ${\cal G}$ are the fully-packed loop configurations just defined.
Interestingly, this model is critical for any
$0 \le |n_{\rm b}|,|n_{\rm g}| \le 2$.

In the Coulomb gases obtained from loop models \cite{ref2} the conditions
defined above are not satisfied in general. In particular, the components
of the matrix $A_{ab}$, cf.~Eq.~(\ref{Cartan}), are not integer-valued for
generic values of $n_{\rm b}$ and $n_{\rm g}$ belonging to the critical
manifold.\footnote{It should be remarked that in the Coulomb gas models derived from the
models of loops \cite{ref1,ref2}, the normalisation of free fields, of
their action, and of the two-point function, is usually different from the
one that we are using, defined by the two-point function
(\ref{twopoint}). This implies a different definition for scalar products
of vectors, like the screening vectors $\{\vec{\alpha}^{\pm}_{a}\}$,
background charge $2\vec{\alpha}_{0}$, and the electric charges
$\{\vec{\beta}\}$ of the physical operators
$V_{\vec{\beta}}(z,\bar{z})$. This also results in a different formula for
the conformal dimensions of the vertex operators. To compare the formulas
in a general setting, independent of the details of a particular model, the
matrix of elastic constants in the free field action of loop models has to
be first diagonalised and then renormalised appropriately, so that the
two-point functions of free fields take the form of Eq.~(\ref{twopoint}),
or, in any case, take a form which is symmetric with respect to the
components of the free fields $\{\varphi_{a}(z,\bar{z})\}$.} 
As a result, defining the ``would be degenerate representations''
and the Kac table would not make sense: the operator algebra of vertex
operators $\{V_{\vec{\beta}_{(...)}}\}$ would not close.  

In these models one defines physical operators differently, by using
physical arguments and constructions that are directly linked to quantities
in the discrete model of oriented loops. The operators coupling to the
electric part%
\footnote{Operators with magnetic Coulomb charge are also physically relevant.
They correspond to topological defects, vortices, in the interfacial
representation, and they are reminiscent of disorder operators attached
to a defect line.}
of the Coulomb gas again take the form of vertex
operators $\{V_{\vec{\beta}}(z,\bar{z})\}$, with $\vec{\beta}$ belonging
to a particular lattice, which e.g.~in case of the two-flavoured loop model
will be a three-dimensional body-centered cubic lattice.
The operator algebra of the operators $\{V_{\vec{\beta}}(z,\bar{z})\}$ 
is closed, by construction. But one will, in general, have to admit an infinite
OPE for a couple of operators $V_{\vec{\beta}_{1}}V_{\vec{\beta}_{2}}$.
As it stands, a model with such properties will not have an extended chiral
algebra, will not have extended symmetries.
It is just a Coulomb gas endowed with conformal invariance, nothing more.

On the other hand, one can check that in the special case
$n_{\rm b}=n_{\rm g}$ of the above-mentioned model the matrix
$A_{ab}$ appearing in Eq.~(\ref{Cartan}) actually {\em does} have
integer valued components:
\beq
 A_{ab}=\frac{2(\vec{\alpha}_{a},\vec{\alpha}_{b})}{(\vec{\alpha}_{b})^{2}}=
 \left(\begin{array}{ccc}2&1&0\\1&2&1\\0&1&2\end{array}\right).
 \label{fakeCartan}
\eeq
The relation (\ref{Cartan1}) between $2\vec{\alpha}_{0}$ and the screenings
is also nicely satisfied, with $m_{a}=(1,0,1)$. The physical operators of
this model (the electric ones) coincide with $\{V_{\vec{\beta}_{(\cdots)}}\}$,
and they satisfy the necessary condition for the degeneracy of the modules in
the way it has been described above. (Incidentally, in the loop model language,
for $n_{\rm b}=n_{\rm g}$ the general model can be rewritten in a way so
that all three elasticity constants are equal \cite{ref3}.)

So far, so good. The worrying point is that the matrix $A_{ab}$ in
Eq.~(\ref{fakeCartan}) does not correspond to a Cartan matrix of any
classical Lie algebra.
The closest one would be the algebra A${}_{3}$, with the Cartan matrix of
the form
\beq
 A_{ab}=\left(\begin{array}{ccc}2&-1&0\\-1&2&-1\\0&-1&2\end{array}\right).
 \label{trueCartan}
\eeq
The fact that the off-diagonal elements of $A_{ab}$ are now negative
is a general property of the Cartan matrix associated with a classical
Lie algebra, it being the matrix of scalar products of simple root vectors.
On the other hand, the condition for the closure of the
algebra of operators $\{V_{\vec{\beta}_{(...)}}\}$ requires, in its general
form presented above, the integer-valuedness only.

\begin{figure}
\begin{center}
 \leavevmode
 \epsfysize=200pt{\epsffile{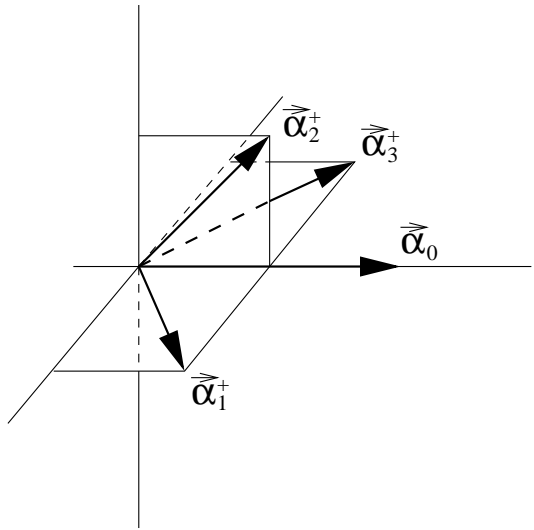}}
 \end{center}
 \protect\caption[3]{\label{fig6}Screening geometry in the two-flavoured
 loop model.}
\end{figure}

\begin{figure}
\begin{center}
 \leavevmode
 \epsfysize=200pt{\epsffile{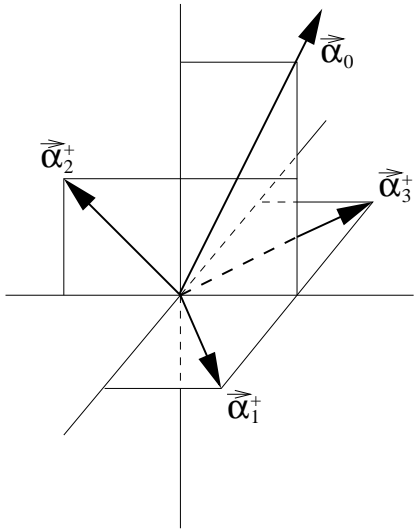}}
 \end{center}
 \protect\caption[3]{\label{fig7}Screening geometry in the WA${}_3$ model.}
\end{figure}

To make more clear the difference between the two sets of vectors
$\{\vec{\alpha}^{+}_{a}\}$, corresponding to the matrices
(\ref{fakeCartan}) and (\ref{trueCartan}), we present them in
Figs.~\ref{fig6} and \ref{fig7}.  Evidently, the vectors in
Fig.~\ref{fig6}, cf.~Eq.(\ref{fakeCartan}), belong to the same lattice as
the vectors in Fig.~\ref{fig7}, but they are not {\em simple} roots of
A${}_{3}$. The vector $\vec{\alpha}^{+}_{2}$ has been switched from one side
to another, leading to a sign change in the off-diagonal elements of the
matrix $A_{ab}$.

It might be possible that a more refined analysis could be given to the
selection of screenings in the Coulomb gas models, taking into account the
properties of representations: Detailed properties of the operator algebra
and of the correlation functions of physical operators. The purpose of the
analysis would be to decide on the acceptance of the theories with the
non-classical matrices $A_{ab}$, as in Eq.~(\ref{fakeCartan}), which appear
in the loop models.

In the absence of such a more detailed analysis, but also to test one more
detail of our approach, we have directly calculated the chiral algebra
operators, the $W$'s, for the theories with the matrices $A_{ab}$ in
(\ref{fakeCartan}) and in (\ref{trueCartan}).  

This last part of our approach is in some sense akin to the inverse
scattering problem: with the data of scattering one has to reconstruct the
Hamiltonian of the problem.  Here we have constructed the Coulomb gases
starting from the geometry of the screening operators. But finally, having
fixed them, it is then perfectly possible to calculate the extended chiral
algebra operators. In particular, in this way it is possible to give a
definitive answer whether an extended chiral algebra exists in the loop
model with the matrix of screenings given by Eq.~(\ref{fakeCartan}).

This last part of the analysis is described and applied to a variety of
particular Coulomb gas problems in the Section which follows. For the loop
model with screening matrix (\ref{fakeCartan}) the answer will be negative.
The model does not have an extended chiral algebra, and its symmetry is
only a conformal one.

In the last Section, which will be devoted to discussion and conclusions,
we shall discuss the possibilities to define, in the loop models setting,
models which might lead to the Coulomb gases related to the classical Lie
algebras, with screenings realised by simple roots. This appears to be the
only possibility to provide them with extended symmetries.

\section{Applications} 

We have seen in the preceding Section that the general consistency
requirements means that the screening vectors must span the root lattice of
a classical Lie algebra. In particular, we would like to consider
situations where this is the case, without actually taking the screening
vectors to be the {\em simple} roots of the concerned algebra. As we have
seen above, one such situation is of interest to the two-flavoured loop
model of Jacobsen and Kondev.

We shall consider various such examples with $D=2$ and $D=3$. It should be
noticed that the fully-packed loop model on the honeycomb lattice (which has
$D=2$) possesses a unique screening vector, which is proportional to the
background charge \cite{ref2}. We have argued that such a situation is
non-generic, and leads to the decoupling of one of the scalar fields from the
other {\em and} from the background charge. Thus, from the point of view of
CFT this theory is trivial. Notwithstanding this ``triviality'', some of the
critical exponents related to geometrical properties of the loops do take
non-trivial values \cite{ref2}, different from those of the standard $D=1$
loop model describing the dense phase of the O($n$) model \cite{ref1}. This is
possible because these exponents do not belong to the physical part of the Kac
table for minimal models. We also point out that despite of the triviality of
its CFT, the model actually has an interesting integrable structure, due to
its underlying $sl_q(2)$ quantum group symmetry \cite{Resh}.

Discarding such trivial possibilities, for both $D=2$ and $D=3$ there are
two different choices of $D$ screening vectors of equal length, spanning
the root lattice of A${}_D$. After defining the method of computations we
turn to the detailed application to those two cases. One of the cases, of
course, is the standard WA${}_D$ geometry, and it serves as a check of our
computations that in this case we reproduce various known results.

\subsection{Definitions}

As already explained, we consider the case of a $D$ dimensional Coulomb gas
with a fixed background charge $\vec{\alpha}_0$ and $D$ screening charges. In
addition to $T$, we must construct $D-1$ extra chiral operators with integer
dimensions. These operators are made as linear combinations of products of
derivatives of free fields. We shall choose here the simplest construction,
namely to search for chiral operators $W_3, \ldots, W_{D+1}$, where the
subscript in $W_i$ indicates its dimension. It is of course possible that the
fields $W_i$ should be found on higher levels, but for reason of simplicity we
find this unlikely. After building the most general operators on the
respective levels, we will impose on them that they commute with the
integrated screening operators, as explained in Section~\ref{sec:kacf}, and
that they be primary operators. The primarity condition is requested in order
to ensure that $\{T(z),W_3(z),\cdots,W_{D+1}(z)\}$ forms a closed algebra.

\begin{figure}
\vskip -7cm
\begin{center}
 \leavevmode
 \epsfysize=190pt{\epsffile{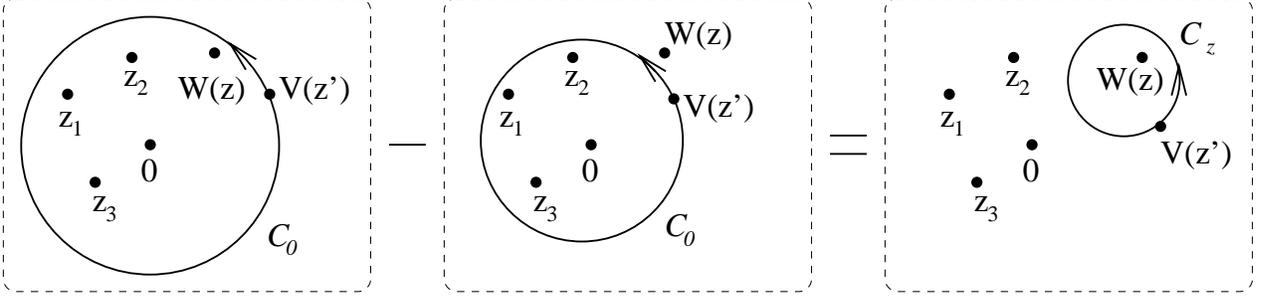}}
 \end{center}
\vskip 5cm 
 \protect\caption[3]{\label{fig8} Deformation of the integration contour in
(\ref{contint}).}
\end{figure}

The requirement that a field $W$ commutes with the
currents of Eq.~(\ref{commute}) can be written 
\beq
\label{contint}
 [W(z),Q]= [W(z),\oint_{C_0} {\rm d}z' \, V(z')] =
  \oint_{C_z} {\rm d}z' \, V(z') W(z) = 0,
\eeq
where we have deformed the integration contour as indicated in
Fig.~\ref{fig8}. This implies that the residue must be zero:
\beq
 {\rm Res}( V(z') W(z) ; z)  = 0.
 \label{residue}
\eeq
Second, $W$ must be a primary operator of dimension $\Delta_W$:
\beq
 \label{Wprimary}
 T(z) W(z') = {\Delta_W \over (z-z')^2} W(z') +  {1 \over (z-z')} \partial
 W(z') +   \cdots.
\eeq
We shall see below that for a specified choice of the screening charges,
these two constraints, when properly expressed, suffice to uniquely determine
the $W$ operator, up to a global normalisation. To fix the latter,
we impose a third constraint:
\beq
\label{Wnormalize}
 W(z) W(z') = {c/\Delta_w \over (z-z')^{2 \Delta_W}} + \cdots.
\eeq

Keeping only the holomorphic part in (\ref{stress}), (\ref{twopoint}) and
(\ref{vertex}), we have 
\bea
T(z)&=& -{1\over 4} \sum_{i=1}^D (\partial \varphi_i)^2 + i \vec{\alpha_0}
\cdot \partial^2 \vec{\varphi} \\
 \langle \varphi_i(z) \varphi_j(z') \rangle &=&  - 2 \delta_{ij} \log(z-z') \,
\eea
with the corresponding central charge:
\beq
c=D - 24 \sum_{i=1}^D (\alpha_0^i)^2
\eeq
and the screening operators:
\beq
V_k(z) = \exp \left(i \sum_i \alpha_k^i \varphi_i(z) \right) 
       = \exp \left( {1\over 2}\sum_i g_k^i \varphi_i(z) \right)
\eeq
with $g_k^i= 2 i \alpha_k^i$. With this last definition, the Wick
contraction with a derivative of a field reads simply
\beq
V_k(z') \partial^i \varphi_j(z) =  {a_i g_k^j  \over (z'-z)^i} V(z') 
\eeq
with $a_1=a_2=1 , a_3=2$ and $a_4=6$. 
Forming suitable linear combinations, this formula will allow us to
compute (\ref{residue}) for any operator $W(z)$ built out of the
fields $\varphi_i(z)$ and their derivatives. However,
we still need to develop the screening operator around $z$: 
\bea
V_k(z') &=& \exp{(\sum_i {g_k^i\over 2} \varphi_i(z'))}   \\
&=& \exp{\left(\sum_i {g_k^i\over 2} (\varphi_i(z) + (z'-z) \varphi'_i
(z) + {(z'-z)^2\over 2} \varphi''_i(z)  + {(z'-z)^3\over 6} 
\varphi'''_i(z) + \cdots)\right)}    \nn \\
&=& V_k(z) \left( 1 + (z'-z) \sum_i {g_k^i\over 2} \varphi'_i(z) 
+ (z'-z)^2 \left[\sum_i {g_k^i\over 4} \varphi''_i(z) + \sum_{i,j}
{g_k^i g_k^j \over 8} \varphi'_i(z)  \varphi'_j(z)\right]
\right. \nn \\
&& \hskip -0.5cm + (z'-z)^3 \left[\sum_i {g_k^i\over 12} \varphi'''_i(z) +
\sum_{i,j} {g_k^i g_k^j \over 8} \varphi'_i(z) 
\varphi''_j(z) \left. + \sum_{i,j,k} {g_k^i g_k^j g_k^l \over 48}
\varphi'_i(z) \varphi'_j(z)  \varphi'_l(z)\right] + \cdots \right), \nn  
\eea
where we have used the notation $\varphi'_i(z')=\partial \varphi_i(z')$,
$\varphi''_i(z')=\partial^2 \varphi_i(z')$ and $\varphi'''_i(z')=\partial^3 
\varphi_i(z')$. 

We consider an operator $W_N$ of dimension $N$, built as a linear combination
of the derivatives of the $D$ scalar fields. Let us define by ${\cal N}_D(N)$
the number of terms in the linear combination which forms the most general operator of dimension $N$.
We then impose the constraint that this operator commutes with the $D$
screening charges. As shown above, this corresponds to canceling the residue in
$z$ of the product $V(z') W_N(z)$, which for a general operator $W_N$ amounts
to canceling an operator of dimension $N-1$. We then expect $D \times {\cal
N}_D(N-1)$ constraints. Next we have to impose the primarity. In the product
$T(z) W_N(z')$, we have to cancel all the powers in $(z-z')^{-i}$ for $i=N+2,
N+1, \ldots, 3$. Since the power $(z-z')^{-i}$ comes with an operator of
dimension $N+2-i$, we will get an additional number of constraints equal to
${\cal N}_D(0)+{\cal N}_D(1)+ \cdots+ {\cal N}_D(N-1)$. Thus for a $W_n$
operator with ${\cal N}_D(N)$ parameters, we have $(D+1) {\cal N}_D(N-1) +
{\cal N}_D(N-2) + \cdots + {\cal N}_D(0)$ constraints. Of course, this is just
the maximal number of constraints that we can expect, since in general not
all of the corresponding linear equations will be independent.

\subsection{Two scalar fields}

The most general operator of dimension three built from two scalar fields is:
\bea
 \label{w3}
 W_3 &=& \sum_i a_i \partial^3 \varphi_i + \sum_i b_i \partial \varphi_i
 \partial^2 \varphi_i + \sum_{i} c_i (\partial \varphi_i)^3 +
 \sum_{i \neq j} d_{ij} (\partial^2 \varphi_i) \partial \varphi_j +
 \sum_{i \neq j} e_{ij} (\partial \varphi_i)^2 \partial \varphi_j. \nn \\ &~&
\eea

We shall choose the coordinate system so that the background electric 
charge is directed along the 1-direction. The stress-energy tensor then
reads
\beq
 T(z) = -\frac14(\partial \varphi_1)^2 - \frac14(\partial \varphi_2)^2
        + i \alpha_0 \partial^2 \varphi_1.
\eeq

For this case, we have ${\cal N}_2(0)=1$, ${\cal N}_2(1)=2$, and ${\cal
N}_2(2)=5$. Thus we will have $8$ constraints from the primarity of $W_3$ and
$10$ constraints from the commutation of the $W_3$ operator with the two
screening operators.

By applying first the constraints from the primarity, we obtain seven
equations on the $10$ constants entering in (\ref{w3}) :
\bea
 &~& \begin{array}{llll}
 a_1    = -4 \alpha_0^2 c_1 &
 b_1    = -6 i \alpha_0 c_1 &
 b_2    = 6 i \alpha_0 (1-8\alpha_0^2) c_1 \\
 e_{21} = -3(1-8 \alpha_0^2) c_1 &
 e_{12} = \frac{i}{4 \alpha_0} (d_{12} + d_{21}) &
 a_2    = -\frac23 i \alpha_0 d_{12} &
 \end{array} \nonumber \\
 &~& 12 i \alpha_0 c_2    = (1-24\alpha_0^2) d_{12} + (1+8 \alpha_0^2)
d_{21}. 
\eea

\subsubsection{Classical WA${}_2$ geometry}

Turning now to the commutation of $W_3$ with the screening operators,
we begin by considering the geometry of the classical WA${}_2$ theory:
\beq
 \vec{\alpha}_1^+ = \alpha_+ \left( \frac12 , \frac{\sqrt{3}}{2} \right)
 \ \ \ \ \ \ 
 \vec{\alpha}_2^+ = \alpha_+ \left( \frac12 , -\frac{\sqrt{3}}{2} \right),
\eeq
where
$\alpha^+$ is related to $\alpha_0$ by the relation (\ref{marginal}), i.e.
$\alpha_0 = {1\over 2}(\alpha^+ - {1\over \alpha^+})$.

The two screenings are related by a reflectional symmetry, and it is thus
no surprise that considering just $\vec{\alpha}_1^+$ leads to a
single-parameter solution
\beq
 \begin{array}{lllll}
 a_1 = 0 &
 a_2 = 24 \alpha_0^2 \tilde{c} &
 b_1 = 0 &
 b_2 = 0 &
 c_1 = 0 \\
 c_2 = 4 \tilde{c} &
q d_{12} = 12 i \alpha_0 \tilde{c} &
 d_{21} = 36 i \alpha_0 \tilde{c} &
 e_{12} = -12 \tilde{c} &
 e_{21} = 0 \; .
 \end{array}
\eeq
If we also impose the normalisation condition (\ref{Wnormalize}), then we
end up with the following expression of the parameters in terms of
the central charge ($c=2-24 \alpha_0^2$): 
\bea
 a_2 &=& {i (2-c)\over 12 \sqrt{5 c +22}} \; \; ; \; \; 
 c_2 = {i \over 3 \sqrt{5 c +22}}  \; \; ; \; \; 
 d_{12} = \pm \sqrt{ (2-c)\over 24 (5 c +22)} \\ 
 d_{21} &=& 3 d_{12} =  \pm \sqrt{ 3(2-c)\over 8 (5 c +22)} \; \; ; \; \; 
 e_{12} = -{i \over \sqrt{5 c +22}} \; .
\eea
Thus, our final result reads
\bea
 T &=& -\frac14(\partial \varphi_1)^2 - \frac14(\partial \varphi_2)^2
        + i \sqrt{2-c \over 24} \partial^2 \varphi_1 \\
W_3 &=&  {1\over \sqrt{5 c +22}}  \left(
{i (2-c)\over 12 } \partial^3 \varphi_2 + {i\over 3} (\partial
\varphi_2)^3 \pm \sqrt{2-c\over 24}  (\partial^2 \varphi_1) \partial
\varphi_2 \right. \nn \\
& & \left. \hskip 2cm \pm \sqrt{3(2-c)\over 8}  (\partial^2 \varphi_2) \partial
\varphi_1 -i (\partial \varphi_1)^2 \partial \varphi_2 \right).
\eea

Up to the normalization constant, this coincides with the solution of
Fateev and Zamolodchikov \cite{ref8}.

\subsubsection{Alternative geometry}
Consider next the alternative geometry of the screening charges
\beq
 \vec{\alpha}_1^+ = \alpha_+ \left( \frac{\sqrt{3}}{2} , \frac12 \right)
 \ \ \ \ \ \ 
 \vec{\alpha}_2^+ = \alpha_+ \left( \frac{\sqrt{3}}{2} , -\frac12 \right).
\eeq
These vectors still span a triangular lattice, i.e.~the root lattice
of the Lie algebra A${}_2$. They also correspond to a ``Cartan matrix''
where the signs of the off-diagonal elements have been changed:
\beq
 \frac{2 (\vec{\alpha}_i,\vec{\alpha}_j)}{(\vec{\alpha}_j)^2} \equiv A_{ij} =
 \left( \begin{array}{cc} 2 & 1 \\ 1 & 2 \end{array} \right).
\eeq
In this case we find only the trivial (null) solution for $W$.

\subsection{Three scalar fields} 

In this section we have found it convenient to impose
the following parametrisation for the screening operators 
\beq
g_1^i = 2 i a (1,0,-1) \; ; \; g_2^i = 2 i a (-1,1,0) \; ; \;
g_3^i = 2 i a (1,0,1)  
\eeq
and for the stress-energy tensor
\beq
\alpha_0^1 = X/2 \; ; \; \alpha_0^2 = X \; ; \; \alpha_0^3 = 0.
\eeq
Here the variable $X=(2 a -1/a)$ is related to the central charge by
$c=3-30 X^2$.

We now have to consider the construction of two operators of
dimensions 3 and 4. 
The most general such operators built from three scalar fields read
\bea
\label{eqw3}
W_3 &=& \sum_i a_i \partial^3 \varphi_i + \sum_i b_i \partial \varphi_i
\partial^2 \varphi_i + \sum_{i} c_i (\partial \varphi_i)^3 
+ \sum_{i \neq j} d_{ij} (\partial^2 \varphi_i) \partial \varphi_j  \nn\\ 
&& + 
\sum_{i \neq j} e_{ij} (\partial \varphi_i)^2 \partial \varphi_j + f \partial
\varphi_1 \partial \varphi_2 \partial \varphi_3
\eea
and
\bea
\label{eqw4}
W_4 &=& \sum_i a_i \partial^4 \varphi_i + \sum_i b_i \partial \varphi_i
\partial^3 \varphi_i + \sum_{i\neq j} c_{ij} \partial^3 \varphi_i \partial
+ \varphi_j \sum_i d_i (\partial^2 \varphi_i)^2 \\ 
&+&  \sum_{i < j} e_{ij} \partial^2 
\varphi_i \partial^2 \varphi_j + \sum_i f_i (\partial^2 \varphi_i)
(\partial \varphi_i)^2 
+ \sum_{i \neq j} g_{ij} (\partial^2 \varphi_i) (\partial \varphi_i)
(\partial \varphi_j) 
\nn\\ 
&+& \sum_{i \neq j} h_{ij} (\partial^2 \varphi_i)
(\partial \varphi_j)^2 + \sum_{i\neq j,k; j<k} i_{ijk} (\partial^2
\varphi_i) (\partial \varphi_j) (\partial \varphi_k)  
+  \sum_i j_i (\partial \varphi_i)^4 \nn \\
&+& \sum_{i \neq j} k_{ij} (\partial
\varphi_i)^3 (\partial \varphi_j) 
+ \sum_{i < j} l_{ij} (\partial \varphi_i)^2 (\partial \varphi_j)^2
+ \sum_{i\neq j,k; j<k} m_{ijk} (\partial \varphi_i)^2 (\partial
\varphi_j) (\partial \varphi_k). \nn
\eea
Note that there is some overlap among the symbols used to designate the
constants entering in the definitions of $W_3$ and $W_4$. These constants
are of course independent in the two cases, but in order to avoid complicating
the notation we do not distinguish them by an extra index.
This convention should lead to no confusion since we shall consider
the operators $W_3$ and $W_4$ separately in the following.

\subsubsection{Classical WA${}_3$ geometry}

In this case we have ${\cal N}_3(0)=1$, ${\cal N}_3(1)=3$, ${\cal N}_3(2)=9$,
${\cal N}_3(3)=22$ and ${\cal N}_3(4)=51$. Thus, Eq.~(\ref{eqw3}) for $W_3$
contains 22 free parameters. Imposing the commutation of $W_3$ with the
screening charges produces $3\times 9$ constraints on these parameters.

Solving the corresponding equations gives the following relations,
leaving just two free parameters, $b_1$ and $d_{13}$ :
\bea
a_1 = - i X b_1 \; \; ; \; \; a_2 &=& -2 i X b_1 \; \; ; \; \;
a_3 = - i d_{13} X \\
b_3 = b_2 = b_1 \; \; ; \; \; d_{32} &=& d_{31} = d_{13}  \; \; ; \; \;
f = i {d13 \over X} .
\eea

Next we impose the primarity of $W_3$, yielding ${\cal N}_3(0)+ {\cal N}_3(1)+
{\cal N}_3(2)=13$ additional constraints. It turns out that the only additional
condition which result from these constraints is $b_1=0$. Thus we end with the
following result :
\bea
T(z)&=& -{1\over 4} \sum_{i=1}^3 (\partial \varphi_1)^2 + i {\sqrt{3-c \over
120}} (\partial^2 \varphi_1  + 2 \partial^2 \varphi_2)  \\   
W_3(z) &\propto& ({3-c\over 30} \partial^3 \varphi_3 + i \sqrt{3-c\over 30}
\left((\partial^2 \varphi_1) 
\partial \varphi_3 + (\partial^2 \varphi_3) \partial \varphi_1 +
(\partial^2 \varphi_3) \partial \varphi_2\right)  - \partial \varphi_1
\partial \varphi_2 \partial \varphi_3). \nn 
\eea
To normalise this properly, we demand that $W(z)W(z') = \frac{c/3}{(z-z')^6} +
\cdots$. It follows that the above solution should be divided by
$-\frac45(7+c)$. The singularity at $c=-7$ will reappear in $W_4$ (see below).


The computation for the $W_4$ operator goes along the same line as for the
$W_3$ case, but with much more parameters and constraints. In this case, we
have ${\cal N}_3(4)=51$ parameters in the definition of $W_4$ (see
(\ref{eqw4})). There are $3 \times {\cal N}_3(3)=66$ constraints produced by
the commutation of the screening operators with $W_4$, and ${\cal N}_3(0)+
{\cal N}_3(1)+ {\cal N}_3(2)+ {\cal N}_3(3)=35$ additional constraints coming
from the requirement that $W_4$ be a primary operator. We thus have 101
constraints on the 51 parameters. Solving these constraints is rather
straightforward since they are just linear equations. Our result (up to
a multiplicative factor) can be expressed as a function of the central
charge $c$ (all the terms not present being zero):
\bea
a_1 = i \frac{\sqrt{3-c}(-26+c+2 c^2)}{\sqrt{30}} \; \; &;& \; \; 
a_2 = -i \frac{\sqrt{3-c} (c-2) (c+7)}{\sqrt{30}}\\
b_1 = b_3 = - 2 (c-6) (c+2) \; \; &;& \; \; 
b_2 = 3 (c-2) (c+7)\\
c_{12} = - (c-3) (5 c + 22) \; \; &;& \; \; 
d_1 = \frac{1}{5} (-18-63 c-2 {c^2})\\
d_2 = \frac{1}{10} (c+7) (19 c-102)  \; \; &;& \; \; 
d_3 = -\frac{9}{2} (c-2) (c+7)\\
e_{12} = -\frac{2}{5} (c-3) (9 c - 2) \; \; &;& \; \;
f_1 = - i 8 \sqrt{\frac{6}{5}} \sqrt{3-c} (c+7)\\
f_2 = 2 f_1  \; \; ; \; \;
g_{12} = g_{31} &=& g_{32} = i \sqrt{30} \sqrt{3-c} (5 c +22)\\
h_{12} = i \sqrt{\frac{6}{5}} \sqrt{3-c} (17 c+54) \; \; &;& \; \;
h_{13} = -i 8 \sqrt{\frac{6}{5}} \sqrt{3-c} (c+7) = f1 \\
h_{21} = h_{23} = i \sqrt{\frac{6}{5}} \sqrt{3-c} (9 c -2)  \; \; &;& \; \;
j_1 = j_2 = j_3 = 12 (c+7)\\
l_{12} = l_{13} = l_{23} &=& -3 (17 c +54)
\eea
with the following solution:
\bea
W_4 &=& i\frac{\sqrt{3-c}(-26+c+2 c^2)}{\sqrt{30}}\partial^4 \varphi_1  +
\frac{1}{5} (-18-63 c-2 {c^2}) (\partial^2 \varphi_1)^2  \nn \\
&+& (c+7)\{ -i\frac{\sqrt{3-c}}{\sqrt{30}} [(c-2) \partial^4 \varphi_2  
+ 48 (\partial^2 \varphi_1 (\partial \varphi_1)^2 + 2 \partial^2 \varphi_2
(\partial \varphi_2)^2 
 + \partial^2 \varphi_1 (\partial \varphi_3)^2)] \nn \\
& & \hskip 1.5cm + 3 (c-2) [\partial \varphi_2\partial^3 \varphi_2
-\frac{3}{2} (\partial^2 
\varphi_3)^2] + \frac{1}{10} (19 c-102) (\partial^2 \varphi_2)^2 \}\nn \\
& & \hskip 1.5cm + 12 ((\partial \varphi_1)^4 + (\partial \varphi_2)^4
+(\partial \varphi_3)^4)) \nn \\ 
&+& (17c + 54 )\{ i\sqrt{\frac{6}{5}} \sqrt{3-c} \partial^2 \varphi_1
(\partial \varphi_2)^2 -3 ((\partial \varphi_1)^2 (\partial \varphi_2)^2 +
(\partial \varphi_1)^2 (\partial \varphi_3)^2  + (\partial \varphi_2)^2
(\partial \varphi_3)^2)\} \nn\\ 
&+& (9 c - 2)i\sqrt{3-c} \{-i\frac{2}{5} \sqrt{3-c} \partial^2 \varphi_1
\partial^2 \varphi_2 + \sqrt{\frac{6}{5}}(\partial^2 \varphi_2 (\partial
\varphi_1)^2 
+ \partial^2 \varphi_2 (\partial \varphi_3)^2)\}\nn\\
&+& i(5 c + 22)\sqrt{3-c} \{ - i\sqrt{3-c} \partial^3 \varphi_1 \partial
\varphi_2 + \sqrt{30}( \partial^2 \varphi_1 \partial \varphi_1 
\partial \varphi_2 \nn \\ 
& & \hskip 3cm + \partial^2 \varphi_3 \partial \varphi_3 \partial
\varphi_1  +\partial^2 \varphi_3 \partial \varphi_3 \partial \varphi_2 )\}
\; .
\eea
Finally, we impose the standard normalisation 
$W(z)W(z') = {c/4 \over (z-z')^8} + \cdots$,
which means that all the above should be divided by the factor
\beq
 192 (2 + c) (7 + c) (22 + 5 c) (114 + 7 c).
\eeq
In particular, the $W_4$ operator becomes singular at a set of special values
of the central charge:
\beq
 c = -7 \ \ \ \ \ \
 c = -2 \ \ \ \ \ \
 c = -{22 \over 5} \ \ \ \ \ \
 c = -{114 \over 7}.
\eeq
These singularities are exactly those found by Blumenhagen et al.~\cite{Blume}.
Kausch and Watts \cite{Watts} find in addition the singularities
$c=1/2$, $c=-68/27$ and $c=-24$ of which we see no sign.

\subsubsection{Alternative geometry} 

Finally, we consider the geometry of the two-flavoured loop model
(cf.~Eq.~(\ref{fakeCartan})), for which we use the following parametrisation
for the screening operators
\beq
g_1^i = 2 i a (1,0,-1) \; ; \; g_2^i = 2 i a (1,1,0) \; ; \;
g_3^i = 2 i a (1,0,1)  
\eeq
and for the stress-energy tensor
\beq
\alpha_0^1 = X/2 \; ; \; \alpha_0^2 = 0 \; ; \; \alpha_0^3 = 0 \; .
\eeq
In this case, very tedious computations show that for $W_3$ as well as for
$W_4$, all the parameters are zero under the application of the constraints.

\section{Discussion}
\label{sec:discussion}
In this paper we have discussed the general construction of CFTs based on
Coulomb gases with several bosonic fields. The requirement that the CFT be
consistent has dictated us two physical guiding principles: The existence of
degenerate representations, and the closure condition on the vertex operator
algebra. From these principles we have obtained the classification condition
(\ref{Cartan}) stating that the screening vectors must belong to the root
lattice of a classical Lie algebra. The screenings are however not required to
be {\em simple} roots, and therefore the matrix $A_{ac}$ in Eq.~(\ref{Cartan})
is more general than the Cartan matrix in the theory of W-algebras.

One of the main motivations of our work has been the attempt to identify the
CFT underlying two different model of fully-packed loops (FPL) \cite{ref2}:
The single-flavoured FPL model on the honeycomb lattice and the two-flavoured
FPL model on the square lattice. Both these loop models are known to be
critical for any value of the loop fugacities in the interval $[-2,2]$. In
both cases, the underlying Coulomb gas furnishes exact values of the central
charge, the thermal scaling dimension, and the scaling dimensions of an
infinite set of topological defects linked to the propagation of a set of
strings between two points (``watermelon dimensions''). In the honeycomb case
this information is further supported by a Bethe ansatz solution, the
integrability of the model being assured by its $sl_q(2)$ quantum group
symmetry \cite{Resh}, whereas in the square case no Bethe ansatz solution is
known, except at the point $(n_{\rm b},n_{\rm g})=(2,2)$ \cite{Nienhuis-BA}.

However, without access to the associated CFT, our knowledge of these models
cannot be considered complete. To illustrate this point it is useful to
compare with the finite-temperature O($n$) model on the honeycomb lattice
\cite{ref1}. Its low-temperature critical phase is described by the dense
phase of a (non-fully packed) loop model, the loops being defined in terms of
the diagrammatic expansion of the associated spin model. This loop model is
solvable, both as a Coulomb gas and by Bethe ansatz techniques, and it
furnishes information on critical indices analogous to the FPL cases mentioned
above. On the other hand, when $n=-2\cos(\pi g)$ with $m=g/(1-g)$ a positive
integer ($m\ge 3$) the model is known to coincide with the series of minimal
models of conventional CFT. For these unitary cases a wealth of further
information is available, and many interesting applications become possible.
To mention but one important example, the knowledge of exact operator product
expansions makes it possible to study perturbatively the coupling of quenched
randomness to the local energy density of such models. For the FPL cases one
could imagine addressing the problem of a compact polymer in a random
environment by similar techniques, but to do so knowing the corresponding CFT
becomes indispensable. Another illustration of the complementarity of the two
approaches (Coulomb gas versus CFT) is that the watermelon dimensions, which
have a straightforward interpretation in terms of the loop model, coincide
with CFT operators that are {\em outside} the physical part of the Kac table
\cite{Duplantier}. Conversely, local physical operators in the CFT description
do not in general seem to have an analogue in the loop approach.

In the present work we have argued that the honeycomb FPL model is in fact
trivial from the CFT point of view, since one of the scalar fields decouples
from the other, and from the background charge. On the other hand, we have
showed that the non-trivial two-flavoured FPL model satisfies the
classification condition (\ref{Cartan}). However, we have also found that the
extended chiral operators needed to control the $D-1$ remaining degrees of
freedom (with $D=3$) do not exist, at least not on level 3
and 4 of the identity module. We can of course not rule out the eventuality
that such operators exist at higher levels, but we find it rather unlikely.
                                                                             
To acquire a consistent CFT description of fully-packed loops, it thus seems
to us that the most natural thing to do would be to somehow modify the
definition of the loop models in question, so that the corresponding CFTs
become the classical W-theories, WA${}_2$ and WA${}_3$ respectively. We recall
that in the Coulomb gas formalism, the operator assigning the proper weights
to the loops is a periodic function on the ideal state graph. The lattice to
which screenings must belong is therefore fixed by standard Fourier analysis
\cite{ref2}. One generally assumes that all of the Fourier modes come with
non-zero amplitudes, and the actual screenings are therefore singled out as
being those closest to the background charge vector, since this ensures the
lowest conformal dimension. For loop models with further adjustable parameters
one may however imagine that the amplitude corresponding to the closest
vectors can be made to vanish, in which case the screenings will have to be
chosen from the vectors second-closest to the background charge, and so on.
This is actually what happens in the O($n$) model, where the temperature $T$
acts as an adjustable parameter. By making the obvious (``closest'') choice
for the screening vector, the one-dimensional Coulomb gas describes the dense
(critical) phase, as mentioned above. But by fine-tuning the temperature it is
possible to access another dilute (tricritical) phase, corresponding to the
choice of the next-closest screening \cite{Duplantier,Kondev-private}. This
situation is rather analogous to standard Ginzburg-Landau theory, where
fine-tuning may serve to make the $\varphi^4$ term vanish, thus giving access
to multi-critical behaviour governed by a more general $\varphi^{2n}$ term. In
a certain sense, successive fine-tuning of more and more parameters is
tantamount to augmenting the symmetry of the corresponding critical theory. We
would expect that the CFTs of the two FPL models under consideration may be
turned into the classical WA${}_2$ and WA${}_3$ theories by fine-tuning
suitable extra parameters, thus driving the systems to multi-criticality.

For the moment we do not have any concrete proposal for the construction of
such tunable parameters. The introduction of temperature-like vacancies in FPL
models is known \cite{Jacobsen-Kondev} to introduce a flow towards the dense
phase of the standard O($n$) model \cite{ref1}, so clearly the temperature is
not a suitable parameter for endowing the model with a {\em higher} symmetry.
A more promising possibility would be to progress in analogy with the
ferromagnetic Ising model, which can be driven to tricriticality by
introducing a staggered magnetic field and fine-tuning its strength (in
addition to the critical temperature). For the fully-packed loop models, it is
possible to impose several staggered fields that act so as to distinguish
between the various ideal states \cite{ref2}.

\noindent{\large\bf Acknowledgments}

We would like to thank D.~Bernard, J.~Kondev and A.~Leclair for stimulating
discussions. We are also grateful to the organizers of the programme
``Integrable models in Condensed Matter and Non-Equilibrium Physics'', May
14 - June 11, 2000, CRM, Universit\'e de Montr\'eal. Part of the present
work has been done and presented at this meeting.

\end{document}